\documentclass[prd,aps,preprint,nofootinbib,superscriptaddress]{revtex4}
\usepackage{epsfig}
\usepackage{amsmath}
\input{epsf}

\begin{document}


\vspace*{2cm}
\title{The evolution of the small $x$ gluon TMD}

\author{Jian Zhou}
\affiliation{\normalsize\it School of physics, $\&$ Key Laboratory of Particle Physics and Particle
Irradiation (MOE), Shandong University, Jinan, Shandong 250100, China} \affiliation{\normalsize\it
 Nikhef and Department of Physics and Astronomy, VU University Amsterdam,  De Boelelaan 1081, NL-1081 HV Amsterdam, The Netherlands}

\begin{abstract}
We study the evolution of the small $x$ gluon transverse momentum dependent(TMD) distribution in
the dilute limit. The calculation has been carried out in the Ji-Ma-Yuan scheme using a simple
quark target model. As expected, we find that the resulting small $x$ gluon TMD simultaneously
satisfies both the Collins-Soper(CS) evolution equation
 and the Balitsky-Fadin-Kuraev-Lipatov(BFKL) evolution equation.
 We thus confirmed the earlier finding that the high energy factorization(HEF) and the TMD factorization
 should be jointly employed to resum the different type large logarithms in a process where
 three relevant scales are well separated.
\end{abstract}

\maketitle

\section{Introduction}
QCD factorization theorems are important tools in describing hard processes of the strong
interaction. Among many factorization frameworks, transverse momentum dependent(TMD)
factorization~\cite{Collins:1981uk,Collins:1984kg} in terms of the Collins-Soper evolution
equation~\cite{Collins:1981uk} or high energy
factorization(HEF)~\cite{Catani:1990eg,Collins:1991ty} in terms of the BFKL evolution
equation~\cite{Kuraev:1977fs,Balitsky:1978ic} should be employed when computing an observable that
is sensitive to parton transverse momentum in high energy scatterings.
 These two factorization frameworks are applicable
in the different kinematical regions. For example, in the process of the color neutral scalar
particle production through gluon fusion($gg\rightarrow H$) in proton proton collisions, the HEF is
valid in the kinematical region $S\gg M^2$ where $S$ is the center mass of energy squared and $M$
is the scalar particle mass, while the TMD factorization holds as long as $p_\perp^2 \ll M^2$ with
$p_\perp$ being the produced scalar particle transverse momentum.

Apparently, there exists an overlap region $S\gg M^2 \gg p_\perp^2$ where the both factorization
formalisms can apply. The natural questions that would be asked then are: do they become equivalent
and produce the same result in the overlap region; or are they complementary to each other in a
such region? A recent study~\cite{Mueller:2012uf} suggests that the later one is a right question
in the sense that the large logarithm terms ${\rm ln}\frac{S}{M^2}$ and ${\rm
ln}\frac{M^2}{p_\perp^2}$ show up at higher orders need to be simultaneously resumed in the small
$x$ formalism and the TMD factorization framework respectively. To be more specific, the large
logarithm ${\rm ln}\frac{S}{M^2}$ in the dilute limit can be
 resummed by means of the BFKL equation which governs the rapidity evolution of the
unintegrated gluon distribution appears in the HEF formula, while the large logarithm ${\rm
ln}\frac{M^2}{p_\perp^2}$ can be conveniently taken care by the Collins-Soper equation that
describes the energy dependence of the gluon TMD appears in the TMD factorization cross section.

In Refs.~\cite{Mueller:2012uf,Mueller:2013wwa}, an explicit next to leading order(NLO) calculation
for the scalar particle production process has been performed in the small $x$ formalism(the color
glass condensate effective theory~\cite{McLerran:1993ni,JalilianMarian:1997gr}) which is reduced to
the HEF in the dilute limit. It has been found that the two different type logarithms can be
resummed consistently at the same time.  The critical step of achieving this is to realize that the
the logarithms ${\rm ln}\frac{S}{M^2}$ and ${\rm ln}\frac{M^2}{p_\perp^2}$ receive contributions
from the clearly separated phase space regions of the radiated gluon. This is because the rapidity
divergence comes from the strong rapidity ordering region and the soft gluon is responsible for the
light cone divergence in the double leading logarithm approximation.  These analysis have been
extended to the other processes, including heavy quark pair production, back-to-back di-jet
production in $eA$ and $pA$ collisions, and Mueller-Navelet Dijet
production~\cite{Mueller:2013wwa,Mueller:2015ael}.

Inspired by Refs.~\cite{Mueller:2012uf,Mueller:2013wwa}, we address the same topic from a different
aspect of view in this paper. Specifically, we do not attempt to extract the large logarithm
contributions from the complete NLO result for a cross section. Instead, we assume that the
unintegrated gluon distribution and the gluon TMD are essentially the same object and share the
same operator definition in the overlap region. One then can compute the LO and NLO contributions
to the gluon TMD/unintegrated gluon distribution starting from their operator definition in a
simple quark target model. We observed that two different logarithms ${\rm ln}\frac{1}{x}$ and
${\rm ln}\frac{x^2 \zeta^2}{l_\perp^2}$ show up simultaneously at the NLO, where $l_\perp$ is the
gluon transverse momentum and $\zeta$ is a parameter proportional to the plus momentum of hadron.
When embedding the gluon TMD/unintegrated gluon distribution into a cross section formula, these
two logarithms will be converted into the logarithms ${\rm ln}\frac{S}{M^2}$ and ${\rm
ln}\frac{M^2}{p_\perp^2}$. Furthermore, we found that the dependencies of the gluon
TMD/unintegrated gluon distribution on the logarithms of $x$ and $\zeta$ are controlled by the BFKL
and CS equation, respectively. In addition, it has been confirmed that the BFKL equation and the CS
evolution are driven by radiated gluons from the different phase space regions. As the first step,
we restrict us to the dilute limit and do not take into account any non-linear saturation
effect~\cite{Balitsky:1995ub,Kovchegov:1999yj,JalilianMarian:1997gr} in this work.

The interplay of TMD/spin physics and small $x$ physics is becoming a topical issue in recent
years~\cite{Mueller:2012uf,Mueller:2013wwa,Bartels:1995iu,Boer:2006rj,Dominguez:2010xd,Dominguez:2011wm,Metz:2011wb,Dominguez:2011br,
Kang:2011ni,Kang:2012vm,Kovchegov:2012ga,Kovchegov:2013cva,Kovchegov:2015zha,Akcakaya:2012si,Altinoluk:2014oxa,
Kotko:2015ura,Schafer:2013mza,Schafer:2014zea,Schafer:2014xpa,Zhou:2015ima,Schafer:2013opa,Zhou:2013gsa,Boer:2015pni,Balitsky:2015qba,Marzani:2015oyb,Kovchegov:2015pbl}.
In particular, we notice that the attempts to unify the description of various types evolution of
the gluon TMD have also been made in
Refs.~\cite{Kovchegov:2015zha,Li:1999tq,Balitsky:2015qba,Marzani:2015oyb}. However, the
calculations presented in Refs.~\cite{Kovchegov:2015zha,Li:1999tq,Balitsky:2015qba,Marzani:2015oyb}
are formulated in the very different ways. It is a non-trivial task to compare the current work
with these formalisms. We thus leave this for the future study. We also would like to mention that
the other type joint resummation has been discussed in the
literatures~\cite{Li:1998is,Laenen:2000ij,Li:2012md}.

The rest of the paper is structured as follows: in the next section, we calculate the next to
leading order correction to the gluon TMD in the small $x$ limit and check if the resulting small
$x$ gluon TMD satisfies the both BFKL and CS evolution equations; in the section III, we comment on
the implications of our result and outline the possible extensions/generalizations of the current
work.

\section{The evolution of the small $x$ gluon TMD}
In this section, we calculate the next to leading order correction to the gluon TMD in a simple
quark target model and show that the derived gluon TMD satisfies both the CS equation and the BFKL
equation simultaneously.
 Our starting point is the matrix element definition for the unpolarized gluon TMD/unintegrated gluon distribution~\cite{Collins:1981uk,Mulders:2000sh,Ji:2005nu},
\begin{eqnarray}
x G(x,l_\perp,x\zeta)  =  \int \frac{d y^- d^2 y_{\perp}}{(2\pi)^3
P^+} \, e^{-ixP^+y^- + il_{\perp} \cdot y_\perp} \langle P |
F^{+}_\mu(y^- , y_{\perp} ){\cal L}_{\tilde n}^\dag(y^-,y_\perp)
{\cal L}_{\tilde n}(0,0_\perp) F^{\mu+}(0) |P \rangle \label{decp}
\end{eqnarray}
where ${\cal L}$ is the process dependent gauge link. In this work, we perform the calculation
using a future pointing staple like gauge link,
\begin{eqnarray}
{\cal L}(\infty; y^-,y_\perp)  ={\cal P } \ {\rm exp} \left ( -ig \int^{\infty}_0 d\lambda \tilde n
\cdot A(\lambda \tilde n+y^-,y_\perp)  \right )
\end{eqnarray}
Here $A^\mu=-if_{abc}A^\mu_c$ is the gluon potential in the adjoint representation. In the
Ji-Ma-Yuan scheme~\cite{Ji:2004wu,Ji:2005nu} (or the old Collins-Soper
scheme~\cite{Collins:1981uk}), $\tilde n$ is defined as a slightly off-light-cone vector $\tilde
n=n+\delta^+ p$ where $0< \delta^+\ll 1$ is a small parameter, and $n=(1,0,0_\perp)$,
$p=(0,1,0_\perp)$ are the commonly defined light cone vectors. With a non-light-like vector $\tilde
n$, the gluon TMD depends on a new scalar $\zeta^2=2(P^+)^2/\delta^+$, where $P^+$ is the nucleon
plus momentum. Apart from the $\zeta$ dependence, the gluon TMD also depend on the renormalization
scale $\mu$, that is not discussed here as it is beyond the scope of this work. Note that a soft
factor in the properly defined gluon TMDs is omitted in the above definition. This does not affect
the result of the current work because  the soft factor in the Ji-Ma-Yuan scheme is independent of
the parameter $\zeta$. Consequently, the subtracted TMDs and the un-subtracted TMDs in the
Ji-Ma-Yuan scheme satisfy the same Collins-Soper evolution equation.
\begin{figure}[t]
\begin{center}
\includegraphics[width=5cm]{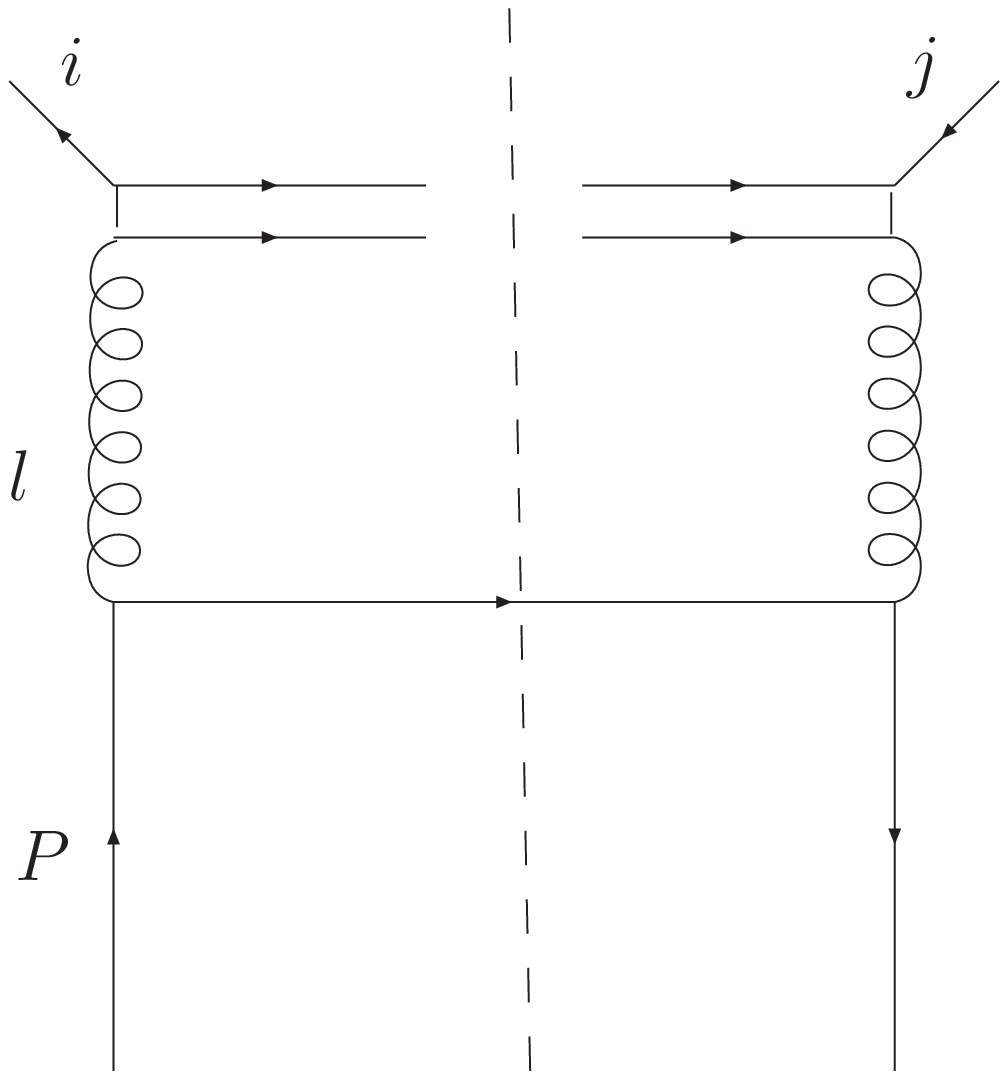}
\caption[] {The leading order diagram contributing to the gluon TMD in the quark target model. $P$
and $l$ are the incoming quark and the outgoing gluon momenta, respectively. $i$, $j$ denote the
gluon polarization indices which will be contracted with $\delta_\perp^{ij}$ in the unpolarized
case.} \label{LO}
\end{center}
\end{figure}

Now we briefly explain why one needs to introduce $\zeta$ dependence by investigating the singularity of the
DGLAP splitting kernel from the real diagrams contributions,
\begin{eqnarray}
{\cal P}_{gg}(z)=2C_A\frac{(z^2-z+1)^2}{z(1-z)}
\end{eqnarray}
where $z=x/x'$ is the longitudinal  momentum fraction of parent gluon carried by the outgoing
gluon. It is easy to see that such splitting kernel is divergent when $z$ approaches 1. This
divergence is often referred to as the light cone divergence which originates from the
contributions of real gluons emitted from gauge links with zero plus momentum. In the case of the
integrated gluon distribution(normal gluon PDF), the light cone  divergence is canceled by the
virtual corrections. However, such cancelation does not occur locally if we leave the gluon
transverse momentum $l_\perp$ unintegrated. A few ways of regulating the light-cone singularity
have been put
forward~\cite{Ji:2004wu,Ji:2005nu,GarciaEchevarria:2011rb,Echevarria:2015uaa,Collins:2011zzd,Li:2014xda},
among which the Ji-Ma-Yuan scheme~\cite{Ji:2004wu} turns out to be the most convenient choice for
the current purpose of this work. In a such scheme, a large logarithm ${\rm ln}
\frac{x^2\zeta^2}{l_\perp^2}$ contribution to the gluon TMD will emerge and can be summed to all
orders utilizing  the CS evolution equation.

On the other hand, it is easy to see that the splitting function are singular when $x$(or $z$) approaches
zero. The large logarithm ${\rm ln}\frac{1}{x}$ contribution  become important in this kinematical limit.
Resummation of this large logarithms in the linear approximation(or in the dilute limit) is accomplished by
the BFKL evolution equation.

In the present paper, we show that the resummation of the two different type large logarithms can
be done in an unified framework. To this end, we first compute the leading order contribution to
the gluon TMD in a simple quark target model.  We then proceed to study how the gluon TMD is
dressed by the quantum corrections at the next to leading order. With the computed gluon TMD at
NLO, we will verify that it satisfies the both CS evolution equation and  BFKL evolution equation.

In the quark target model, the leading order contribution to the gluon TMD only comes from one
graph illustrated in the Fig.\ref{LO} and is given by,
\begin{eqnarray}
 G(x,l_\perp,x\zeta)_{LO}=\frac{\alpha_s C_F}{2\pi^2}
\frac{1}{l_\perp^2}  \frac{1+(1-x)^2}{x}
\end{eqnarray}
where $x=l^+/P^+$ with $P^+$ and $l^+$ being the incoming quark and the outgoing gluon momenta respectively.
In the small $x$ limit, it is simplified as,
\begin{eqnarray}
x G(x,l_\perp,x\zeta)_{LO}|_{x\rightarrow 0} =
 \frac{\alpha_s C_F}{\pi^2} \frac{1}{l_\perp^2}
 \label{lo}
\end{eqnarray}
Here one immediately notices that the expression for the gluon TMD at the leading order contains
neither the logarithm ${\rm ln} \frac{1}{x}$ nor the logarithm ${\rm ln}
\frac{x^2\zeta^2}{l_\perp^2}$. It is thus necessary to extend the analysis to the next to leading
order for the purpose of studying the evolution of small $x$ gluon TMD.

\subsection{Real corrections}
\begin{figure}[t]
\begin{center}
\includegraphics[width=16cm]{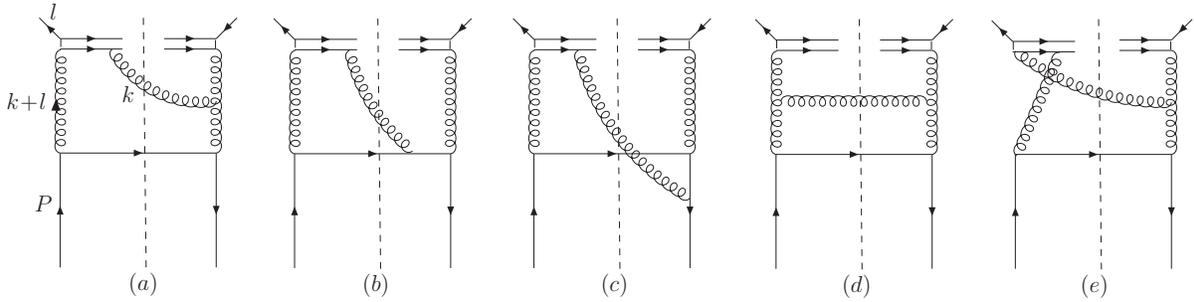}
\caption[] {Sample real diagrams contributing to the evolution kernels of the CS equation and the BFKL
equation. All real diagrams shown here  contribute to the evolution kernel of the BFKL equation. The Fig.(a)
and its conjugate diagram are the only real corrections that contribute to both the evolution kernels of the
BFKL equation and the CS equation. } \label{realf}
\end{center}
\end{figure}

In this subsection, we calculate the NLO real corrections to the gluon TMD. At NLO,  there are a
number of diagrams contributing to the gluon TMD, some of which have been shown in the Fig.2. Among
these real graphs, the Fig.\ref{realf}(a) is a particularly interesting one, because the
Fig.\ref{realf}(a) and its conjugate diagrams are the only real diagrams contributing to the both
evolution kernels of the BFKL and the CS equations. It would be instructive to present some
technical steps for the calculation of the Fig.\ref{realf}(a).

The calculation of the real correction from the Fig.\ref{realf}(a) involves the following integration,
\begin{eqnarray}
I_{\ref{realf}(a)}&=&{\cal C} \int dl^- d^4k \delta \left ((P-l-k)^2 \right )
\delta(k^2)\theta(k^+) \theta(P^+-l^+-k^+) \nonumber \\ && \times
 \frac{2(P^+-l^+-k^+) [ k^+ (k_\perp+l_\perp)\cdot l_\perp
-2l^+(l_\perp+k_\perp)^2  ]
 }{[(k+l)^2+i \epsilon]^2[k^++\delta^+ k^-+i \epsilon][l^2+i\epsilon]}
\end{eqnarray}
where ${\cal C}=\alpha_s^2 (N_c^2-1)/4\pi^4$.  The radiated gluon momentum is denoted as $k$. The
first step is to carry out the integration over $l^-$ and $k^-$ using the delta function which
comes from the on shell condition.
 After integrating out $l^-$ and $k^-$, one has the kinematical constraints  $l^-+k^- = -(l_\perp+k_\perp)^2/2(P^+-l^+-k^+)$ and
$k^-=k_\perp^2/2k^+$. To simplify the calculation, we make the approximation $-l^- \approx k^- =
\frac{k_\perp^2}{2k^+} $ that is valid in the small $x$ region, and obtain,
\begin{eqnarray}
I_{\ref{realf}(a)} = - {\cal C} \int d^2k_\perp \int_0^{P^+}  d k^+
 \frac{k^+ [ k^+ (k_\perp+l_\perp)\cdot l_\perp
-2l^+(l_\perp+k_\perp)^2  ] }{(k_\perp+l_\perp)^4 [2(k^+)^2+\delta^+
k_\perp^2][l^+k_\perp^2+k^+l_\perp^2]}
\end{eqnarray}

We proceed by separating $k^+$ integration into two parts $\int_0^{P^+} dk^+=\int_0^{l^+} dk^+ +
\int_{l^+}^{P^+} dk^+$. Integrating out $k^+$ and  keeping only the large logarithm terms, one
arrives,
\begin{eqnarray}
I_{\ref{realf}(a)} &=& {\cal C} \int d^2k_\perp
  \frac{ 1} {(k_\perp+l_\perp)^2 2 \left [k_\perp^2+l_\perp^4/x^2\zeta^2 \right] }
  {\rm ln}\frac{k_\perp^2(k_\perp^2+x^2\zeta^2)}{( k_\perp^2+ l_\perp^2) ^2}
 +{\cal O}\left (  \frac{1}{{\rm ln}\frac{x^2\zeta^2}{  l_\perp^2 }} \right ) \nonumber \\ && -{\cal
C} {\rm ln}\frac{ 1 }{ x } \int_0 d^2k_\perp \frac{ (k_\perp+l_\perp)\cdot l_\perp }
{(k_\perp+l_\perp)^4 }
 \frac{1}{2l_\perp^2}
+ {\cal O}\left (  \frac{1}{{\rm ln}\frac{ 1 }{ x }} \right )
\end{eqnarray}
where it is easy to see that two different type logarithms arise from the clearly separated phase
space regions. To be more precise, the Collins-Soper type evolution is driven by the emitted gluon
which carries  very small longitudinal momentum $k^+ \ll l^+$. In contrast, the BFKL  evolution
kernel only receives the contribution from the phase space region where $k^+ \gg l^+$, the
so-called strong rapidity ordering region. This finding is consistent with the observation made in
Ref.~\cite{Mueller:2012uf}.

 \begin{figure}[t]
\begin{center}
\includegraphics[width=15cm]{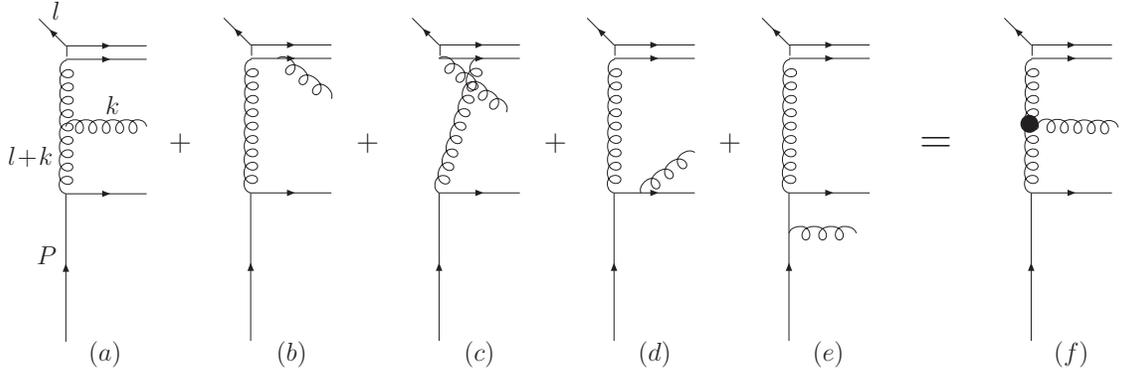}
\caption[] {In the strong rapidity ordering region $l^+ \ll k^+ \ll
P^+$, all real corrections Fig.(a)-Fig.(e) can be summarized into
one diagram Fig.(f) with an effective Lipatov vertex. The large
solid circle denotes the Lipatov vertex.} \label{Lipatov}
\end{center}
\end{figure}

At this step, we have completed the calculation of the real correction to the CS evolution kernel.
However, as mentioned  earlier, many other real diagrams give rise to the large logarithm ${\rm ln}
\frac{1}{x}$ contribution as well.  Instead of calculating them one by one by brute force,  we
simplify the calculation by isolating the leading logarithm contribution with the help of the
so-called  effective Lipatov vertex. It is well known that the leading logarithms of $x$ in the
BFKL dynamics are generated in the strong rapidity ordering region: $P^+ \gg k^+ \gg l^+$. In a
such kinematical region, it is justified to approximate the incoming and outgoing quark lines as
the Eikonal lines. After making this approximation, as illustrated in the Fig.3, graphs with all
possible gluon insertions can be summarized into one diagram with an effective Liaptov vertex. The
effective Lipatov vertex reads,
\begin{eqnarray}
C^\mu=g \left \{ (k_T+2l_T)^\mu-\left [ k^+-\frac{(k_\perp+l_\perp)^2}{k^-} \right ] p^\mu +\left
[k^--\frac{l_\perp^2}{k^+} \right ] n^\mu \right \}
\end{eqnarray}
where $k_T$ is the transverse component of four momentum with $k_T^2=-k_\perp^2$. Such an effective
vertex is gauge invariant and satisfies $ C_\mu k^\mu = 0$.
 By simply squaring the amplitude and using the relation $C^\mu C_\mu =g^2
\frac{-4(k_\perp+l_\perp)^2l_\perp^2}{k_\perp^2} $, it is straightforward to calculate the real
gluon contribution to the logarithm  ${\rm ln} \frac{1}{x}$ terms.  The total real correction to
the gluon TMD in the leading logarithms approximation is then given by,
\begin{eqnarray}
x G(x,l_\perp,x\zeta)_{Rel}|_{x\rightarrow 0}&=& {\cal C} \int d^2k_\perp
  \frac{ 1} {(k_\perp+l_\perp)^2  \left [k_\perp^2+l_\perp^4/x^2\zeta^2 \right] }
  {\rm ln}\frac{k_\perp^2(k_\perp^2+x^2\zeta^2)}{( k_\perp^2+ l_\perp^2) ^2}
  \nonumber \\
 && + 2{\cal C} {\rm ln}\frac{ 1 }{ x } \int_0  \frac{ d^2k_\perp} {(k_\perp+l_\perp)^2 k_\perp^2}
 \label{real}
\end{eqnarray}
Using the Eq.~\ref{LO}, the NLO real correction to the gluon TMD can be reexpressed in terms of the
leading order gluon TMD,
\begin{eqnarray}
x G(x,l_\perp,x\zeta)_{Rel}|_{x\rightarrow 0}=  \frac{\alpha_s N_c}{\pi^2}
 \int \! d^2k_\perp
 \left \{  \frac{{\rm ln}\frac{1}{x}}{k_\perp^2}+
  \frac{ {\rm ln}\frac{k_\perp^2(k_\perp^2+x^2\zeta^2)}{( k_\perp^2+ l_\perp^2) ^2}} {  2\left [k_\perp^2+l_\perp^4/x^2\zeta^2 \right] }
   \right \}
  x G_{LO}(x,k_\perp+l_\perp,x\zeta)
\end{eqnarray}

\subsection{Virtual corrections}
\begin{figure}[t]
\begin{center}
\includegraphics[width=11cm]{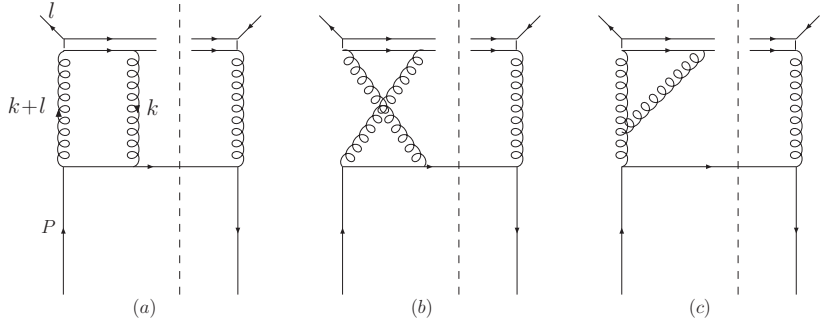}
\caption[] {Virtual diagrams contributing to the gluon TMD at NLO. The Fig.a and Fig.b and their
conjugate diagrams give rise to the virtual corrections to the BFKL evolution kernel, while the
Fig.c and its conjugate diagram contribute to the CS evolution kernel. } \label{virtualf}
\end{center}
\end{figure}

We now move on to evaluate virtual corrections. Among many virtual diagrams, the
Fig.\ref{virtualf}a, Fig.\ref{virtualf}b and their conjugate diagrams produce the large logarithm
of $x$~\cite{yuri,Fleming:2014rea}.  On the other hand,  the light  cone divergence is only
generated from the Fig.\ref{virtualf}c and  its mirror diagram.

Let us first focus on isolating the leading logarithm of  $x$ from the Fig.\ref{virtualf}a, whose
contribution reads
\begin{eqnarray}
I_{\ref{virtualf}(a)}&=&\frac{-i{\cal C}}{4\pi}\int dl^- \int d^4k
\\ \nonumber && \times
\frac{4(P^+-l^+)(P^+-k^+-l^+)[(k_\perp+l_\perp)\cdot l_\perp]}{[(k+l)^2+i \epsilon][k^2+i
\epsilon][(P-k-l)^2+i \epsilon][k^++i \epsilon][l^2+i\epsilon]} \delta((P-l)^2)
\end{eqnarray}
where the Eikonal approximation has been applied to the quark line. In the above formula, we do not
regularize the gauge link propagator as this integral is free from the light cone singularity.
 It is straightforward to carry out $l^-$ integration using the delta function,
\begin{eqnarray}
I_{\ref{virtualf}(a)} &=& \frac{-i{\cal C}}{4\pi}\int d^4k
\frac{2(P^+-k^+-l^+)}{[2(k^++l^+)k^--(k_\perp+l_\perp)^2+i \epsilon] [2k^+k^--k_\perp^2+i
\epsilon]}
 \nonumber \\&& \ \ \ \ \ \ \ \ \ \ \ \ \ \times
\frac{[(k_\perp+l_\perp)\cdot l_\perp]}{[-2(P^+-k^+-l^+)k^--(k_\perp+l_\perp)^2+i \epsilon][k^++i
\epsilon][l_\perp^2]}
\end{eqnarray}
Here and hereafter, we ignore  $l^- \approx -l_\perp^2/2P^+$ that appears in the integral. We then
proceed to do the $k^-$  integration by closing different contours depending on the value of $k^+$,
\begin{eqnarray}
I_{\ref{virtualf}(a)} &= & \frac{{\cal C}}{2} \left \{ \int_{-l^+}^0 dk^+ d^2k_\perp
 \frac{(k_\perp+l_\perp)\cdot l_\perp}{(k_\perp+l_\perp)^2l_\perp^2}
\frac{k^++l^+}{[k^+(k_\perp+l_\perp)^2-(k^++l^+)k_\perp^2+i \epsilon]} \frac{1}{[k^++i \epsilon]}
\right .\
 \nonumber \\ & +&  \left .\ \!\!\!\!
\int_0^{P^+-l^+}  \!\!\! dk^+ d^2k_\perp
 \frac{(k_\perp+l_\perp)\cdot l_\perp}{(k_\perp+l_\perp)^2l_\perp^2}
\frac{P^+-k^+-l^+}{[k^+(k_\perp+l_\perp)^2-(P^+-k^+-l^+)k_\perp^2+i \epsilon]} \frac{1}{[k^++i
\epsilon]} \right \}
\end{eqnarray}
One notices that each of two integrals contains the light cone singularity when $k^+$ approaches
zero. However, the sum of them is free from  the light cone divergence and is given by,
\begin{eqnarray}
I_{\ref{virtualf}(a)} &=&  \frac{{\cal C}}{2}\int d^2 k_\perp  \frac{(k_\perp+l_\perp)\cdot
l_\perp}{(k_\perp+l_\perp)^2l_\perp^2 k_\perp^2}
 \nonumber
\\ &&\times
 \left [{\rm ln}\frac{l^+}{P^+ \!\! -l^+}
+\frac{2k_\perp^2(k_\perp+l_\perp)^2 }{(k_\perp+l_\perp)^4-k_\perp^4}{\rm ln}\frac{k_\perp^2}{(k_\perp+l_\perp)^2}
+i\pi \frac{k_\perp^2+2(k_\perp+l_\perp)^2}{k_\perp^2+(k_\perp+l_\perp)^2}
\right ]
\end{eqnarray}
where the leading logarithm  term can be further re-expressed as,
\begin{eqnarray}
I_{\ref{virtualf}(a)} &=&
 -\frac{{\cal C}}{4}{\rm ln}\frac{1}{x} \int d^2k_\perp \frac{1}{k_\perp^2}\frac{1}{(k_\perp+l_\perp)^2}
 + {\cal O}\left ( \frac{1}{{\rm ln} \frac{1}{x}} \right )
 \label{figva}
\end{eqnarray}
In arriving at the above result, we have made use of  the following identity,
\begin{eqnarray}
\int d^2 k_\perp \frac{(k_\perp+l_\perp) \cdot l_\perp}{k_\perp^2 (k_\perp+l_\perp)^2}= -\int d^2
k_\perp \frac{k_\perp \cdot l_\perp}{k_\perp^2 (k_\perp+l_\perp)^2}
\end{eqnarray}
which can be easily verified by changing the integration variable $k_\perp \rightarrow
-k_\perp-l_\perp$. Repeating the similar procedure,
 for the Fig.\ref{virtualf}b, one obtains,
\begin{eqnarray}
I_{\ref{virtualf}(b)} &=&
 -\frac{{\cal C}}{4} {\rm ln}\frac{1}{x} \int d^2k_\perp \frac{1}{k_\perp^2}\frac{1}{(k_\perp+l_\perp)^2}
+ {\cal O}\left ( \frac{1}{{\rm ln} \frac{1}{x}} \right ) \label{figvb}
\end{eqnarray}

To complete the calculation of virtual corrections in the leading logarithm approximation, we also
need include the contribution from the Fig.\ref{virtualf}(c),
\begin{eqnarray}
I_{\ref{virtualf}(c)}= \frac{i{{\cal C} \mu^{2\epsilon} }}{(2\pi)^{1-2\epsilon}}\int dl^- \int d^{
4-2\epsilon}k \frac{2(P^+-l^+)[2(l^++k^+)l_\perp^2+2 k^+ l_\perp \cdot k_\perp]} {[(k+l)^2+i
\epsilon][k^2+i \epsilon][k^++\delta^+ k^-+i \epsilon][l^2+i\epsilon]^2} \delta((P-l)^2)
\end{eqnarray}
where we use  dimensional regularization to treat ultraviolet divergences. It is  again convenient
to first carry out the integration over $l^-$ and $k^-$. One obtains,
\begin{eqnarray}
I_{\ref{virtualf}(c)} &=& \frac{{\cal C} \mu^{2\epsilon}}{(2\pi)^{-2\epsilon}} \left \{
\int_{-l^+}^0 \!\! dk^+ \! \int \! d^{ 2-2\epsilon} k_\perp \frac{ k^+ \left [
2(l^++k^+)l_\perp^2+2 k^+ l_\perp \cdot k_\perp \right ]}
{[(k^++l^+)k_\perp^2-k^+(k_\perp+l_\perp)^2] [2(k^+)^2+\delta^+ k_\perp^2][l_\perp^4]}
\right .\ \nonumber \\
&& \left .\ + \int_{-\infty}^{-l^+} \!\! dk^+ \! \int d^{ 2-2\epsilon}\! k_\perp \frac{ \delta^+
\left [ 2(l^++k^+)l_\perp^2+2 k^+ l_\perp \cdot k_\perp \right ]} {[2(k^++l^+)k^+
+\delta^+(k_\perp+l_\perp)^2] [2(k^+)^2 + \delta^+ k_\perp^2][l_\perp^4]} \right \}
 \end{eqnarray}
Here $l^-\approx -l_\perp^2/2P^+$ component has been neglected.  It is easy to further integrate
out $k_\perp$ by applying the Feynman parametrization. We are then only left with the integration
with respect to $k^+$ and the Feynman parameter. At this step, though  it is still hard to get the
complete analytical result, one can readily extract the double leading logarithm and the single
leading logarithm contributions,
\begin{eqnarray}
I_{\ref{virtualf}(c)}
 = \frac{{\cal C} \pi}{2 l_\perp^2}\left [ \frac{1}{2\epsilon}-\frac{1}{2} {\rm
 ln}\frac{x^2 \zeta^2}{\mu^2}+
 {\rm ln}\frac{x^2\zeta^2}{l_\perp^2} -\left( {\rm ln}\frac{x^2\zeta^2}{l_\perp^2} \right )^2 \right ]
 +{\cal O}\left( \frac{1}{{\rm ln}\frac{x^2\zeta^2}{l_\perp^2}}\right )
\end{eqnarray}
where the ultraviolet pole can be simply removed according to the minimal subtraction scheme. One
notices that the off-shellness of the incoming gluon serves a natural infrared cut off. At this
point, it is worthy mentioning that the double leading logarithm term in the above formula only
receives the contribution from the phase space region where $|k^+| \ll |l^+|$ which is beyond the
scope of the strong rapidity ordering region. Collecting all contributions from the
Fig.\ref{virtualf}a, Fig.\ref{virtualf}b, Fig.\ref{virtualf}c and their mirror graphs together, the
total virtual correction to the small $x$ gluon TMD is given by,
\begin{eqnarray}
x G(x,l_\perp,x\zeta)_{Vir}|_{x\rightarrow 0}= -{\cal C} \ {\rm ln}\frac{1}{x} \int \frac{
d^2k_\perp}{(k_\perp+l_\perp)^2k_\perp^2} +
 \frac{{\cal C}\pi}{ l_\perp^2}\left [{\rm ln}\frac{x^2\zeta^2}{l_\perp^2} -\frac{1}{2} {\rm
 ln}\frac{x^2 \zeta^2}{\mu^2}-\left( {\rm ln}\frac{x^2\zeta^2}{l_\perp^2} \right )^2 \right ]
 \label{virtual}
\end{eqnarray}
Using Eq.\ref{lo}, it can be re-expressed  as,
\begin{eqnarray}
x G(x,l_\perp,x\zeta)_{NLO}&= &\frac{\alpha_s N_c}{2\pi} \left [ {\rm ln}\frac{x^2
\zeta^2}{l_\perp^2} -\frac{1}{2} {\rm
 ln}\frac{x^2 \zeta^2}{\mu^2} -\left ({\rm ln}\frac{x^2 \zeta^2}{l_\perp^2}\right )^2 \right ] x
G_{LO}(x,l_\perp,x\zeta) \nonumber  \\ &-&
 \frac{\alpha_s N_c}{2\pi^2}{\rm ln}\frac{1}{x} \int \frac{d^2k_\perp}{k_\perp^2}
 \frac{l_\perp^2}{(l_\perp+k_\perp)^2} x
   G_{LO}(x,l_\perp,x\zeta)
\end{eqnarray}

\subsection{The gluon TMD at NLO in the leading logarithms approximation }
Combining the real correction and the virtual correction obtained in the previous subsections, up
to the leading logarithms accuracy the gluon TMD at NLO is given by,
\begin{eqnarray}
x G(x,l_\perp,x\zeta)_{NLO}= x G(x,l_\perp,x\zeta)_{LO}+x G(x,l_\perp,x\zeta)_{Rel}+x
G(x,l_\perp,x\zeta)_{Vir}
\end{eqnarray}
 The gluon TMD at NLO in the small $x$ then reads,
\begin{eqnarray}
x G(x,l_\perp,x\zeta)_{NLO}&= & x G(x,l_\perp,x\zeta)_{LO} \nonumber  \\ &+&
 \frac{\alpha_s N_c}{\pi^2}{\rm ln}\frac{1}{x} \int \frac{d^2k_\perp}{k_\perp^2}
   \left [ x G_{LO}(x,k_\perp+l_\perp,x\zeta)- \frac{l_\perp^2}{2(l_\perp+k_\perp)^2} x
   G_{LO}(x,l_\perp,x\zeta)\right ]
\nonumber  \\ &+&\frac{\alpha_s N_c}{2\pi} \left [ {\rm ln}\frac{x^2 \zeta^2}{l_\perp^2}
-\frac{1}{2} {\rm ln}\frac{x^2 \zeta^2}{\mu^2}-\left ({\rm ln}\frac{x^2 \zeta^2}{l_\perp^2}\right
)^2 \right ] x G_{LO}(x,l_\perp,x\zeta)
\nonumber  \\
&+&\frac{\alpha_s N_c}{\pi^2}\int \frac{d^2k_\perp}{2\left [k_\perp^2+l_\perp^4/x^2\zeta^2 \right]}
  {\rm ln}\frac{k_\perp^2(k_\perp^2+x^2\zeta^2)}{( k_\perp^2+ l_\perp^2) ^2}
 x G_{LO}(x,k_\perp+l_\perp,x\zeta)
\end{eqnarray}
This is the main result of our paper.

We are now ready to check if the computed small $x$ gluon TMD at NLO satisfies the both BFKL and CS
evolution equations. For the ${\rm ln}\frac{1}{x}$ dependence, it is easy to see that the resulting
gluon TMD satisfies the following equation in the strong rapidity ordering region,
\begin{eqnarray}
\frac{\partial \left [ x G(x, l_\perp,x\zeta) \right ]}{\partial {\rm ln}(1/x)} &=&\frac{\alpha_s
N_c}{\pi^2} \int \frac{d^2k_\perp}{k^2_\perp} \left \{ x
G(x,k_\perp+l_\perp,x\zeta)-\frac{l_\perp^2}{2(l_\perp+k_\perp)^2}  x G(x,l_\perp,x\zeta) \right \}
\end{eqnarray}
which is just the famous BFKL evolution equation. For the ${\rm ln}\zeta$ dependence, the
derivative of the computed  gluon TMD with respect to ${\rm ln}\zeta$ is given by,
\begin{eqnarray}
\frac{\partial \left [  G(x, l_\perp,x\zeta) \right ]}{\partial {\rm ln}\zeta} &=&\frac{\alpha_s
N_c}{\pi^2} \int d^2k_\perp  \frac{1}{k_\perp^2+l_\perp^4/x^2\zeta^2} G(x,k_\perp+l_\perp,x\zeta)
\nonumber \\&&
 +\frac{\alpha_s N_c}{\pi}\left [\frac{1}{2}-2 {\rm ln}\frac{x^2 \zeta^2}{l_\perp^2} \right ]
G(x,l_\perp,x\zeta)
\end{eqnarray}
where the terms suppressed by the power of $1/\zeta^2$ have been neglected. In the impact parameter
$b_\perp$ space, it takes a more conventional form,
\begin{eqnarray}
\frac{\partial \left [  G(x, b_\perp,x\zeta) \right ]}{\partial {\rm ln}\zeta} =- \frac{\alpha_s
N_c}{\pi}  \ {\rm ln} \left [ \frac{x^2 \zeta^2 b_\perp^2}{4}e^{2\gamma_E-\frac{1}{2}} \right ]
G(x,b_\perp,x\zeta)
\end{eqnarray}
which can be recognized as the famous CS evolution equation for the gluon TMD. However, our result
differs from that derived in Ref.~\cite{Ji:2005nu} by a finite term.  More work will be needed to
settle down this issue as we are not able to localize the source of this discrepancy for the time
being.  Nevertheless, based on the above findings, it might be fair to claim that it is possible to
describe the BFKL and CS dynamics in an unified framework.

\section{Summary}
We end this paper with a short summary of our major results and a number of remarks on their
implications. We compute the NLO correction to the small $x$ gluon TMD in the leading logarithm
approximation starting from the well known operator definition of the gluon TMD. It is shown that
the resulting gluon TMD simultaneously satisfies the both  CS equation and  BFKL equation. One thus
 may conclude that the unintegrated gluon distribution and the gluon TMD share the same operator
definition in the overlap region where both the TMD factorization and the HEF  apply. Moreover, we
confirmed the observation made in Refs.~\cite{Mueller:2012uf,Mueller:2013wwa} that the BFKL
evolution kernel and the double leading logarithm part of the CS evolution kernel receive
contributions from the clearly separated phase space regions. To be more specific, the double
leading logarithm contribution to the CS evolution is yielded from the phase space region outside
the scope of the strong rapidity ordering region.

In order to resum two different type but equally important logarithms  ${\rm ln}\frac{S}{M^2}$ and
${\rm ln}\frac{M^2}{p_\perp^2}$ in the overlap region which is $S \gg M^2 \gg p_\perp^2$ in the
scalar particle production case, one should employ the TMD factorization and the HEF(or CGC in the
dense medium region) jointly~\cite{Mueller:2012uf,Mueller:2013wwa}. The soft parts in the two
frameworks can be treated in an unified way as shown above. Now we turn to discuss the
perturbatively calculable hard coefficients appears in the TMD factorization and the HEF.
 The impact factor is calculated  with the off shell incoming gluons
in the HEF framework, while the hard part from the TMD factorization is estimated with on shell
gluons.  When $M^2 \gg p_\perp^2$, it is justified to Taylor expand the impact fact in terms of the
power $l_\perp/M$ where $l_\perp$ is the incoming gluon transverse momentum. It is easy to verify
that the non-vanishing leading term of such power expansion is identical to the hard part appears
in the TMD factorization formula using the Ward identity
argument~\cite{Collins:1991ty,Eguchi:2006mc,Collins:2008sg,Liang:2008rf}. The similar equivalence
between the TMD and the CGC has also been established for the dense medium region
case\cite{Dominguez:2010xd,Dominguez:2011wm,Akcakaya:2012si,Kotko:2015ura}.

There are a number of directions in which our work could be extended. First, the natural next step
is to incorporate the saturation
effect~\cite{Balitsky:1995ub,Kovchegov:1999yj,JalilianMarian:1997gr}. Second, it should be feasible
to apply the same analysis to the polarized cases. This is because  the linearly polarized gluon
TMD and the dipole type T-odd gluon TMDs inside a transversely polarized target have been found to
grow very rapidly towards small $x$ as
well~\cite{Metz:2011wb,Dominguez:2011br,Schafer:2013opa,Zhou:2013gsa,Boer:2015pni}. We anticipate
that two different type large logarithms also show up at the same time for these polarization
dependent gluon TMDs in the small $x$ limit. However, we would like to emphasize that the small $x$
evolution of the gluon helicity distribution has to be treated in a complete different
way~\cite{Bartels:1995iu,Kovchegov:2015pbl}. Third, by extracting the large logarithm terms from
the complete two loop order results for TMDs~\cite{Gehrmann:2014yya,Echevarria:2015usa} one should
be able to reproduce the same result presented in this paper. Finally, we have demonstrated that
the Ji-Ma-Yuan scheme is compatible with the small $x$ formalism. Though there is no reason to
doubt that other schemes are not compatible with the small $x$ formalism, it would be nice to check
them through explicit calculations.

\noindent {\bf Acknowledgments:} I thank Feng Yuan, Hsiang-nan Li, Yu-ming Wang,
Miguel~G.~Echevarria, and  Tomas Kasemets for helpful discussions. This research has been supported
by the EU "Ideas" program QWORK (contract 320389).

\end {document}